\documentclass[preprint,12pt]{article}
\usepackage{amssymb}
\usepackage{graphicx,amsmath,amssymb,amsfonts}

\renewcommand{\L}{{\cal L}}

\renewcommand{\v}{\upsilon}

\usepackage{fullpage}

\newcommand{\stab}{{\rm stab}}
\newcommand{\ad}{{\rm ad}}
\newcommand{\EQ}{\begin{equation}}
\newcommand{\EE}{\end{equation}}
\newcommand{\EQA}{\begin{eqnarray}}
\newcommand{\EEA}{\end{eqnarray}}

\def\longrightharpoonup{\relbar\joinrel\rightharpoonup}
\def\longleftharpoondown{\leftharpoondown\joinrel\relbar}
\usepackage{color}   

\def\longrightleftharpoons{
  \mathop{
    \vcenter{
      \hbox{
	\ooalign{
	  \raise1pt\hbox{$\longrightharpoonup\joinrel$}\crcr
	  \lower1pt\hbox{$\longleftharpoondown\joinrel$}
	}
      }
    }
  }
}

\usepackage[numbers, compress]{natbib}
\usepackage[font={small}]{caption}
\begin{document}

\begin{titlepage}
\title{Universality and predictability in molecular\\ quantitative genetics}
\author{Armita Nourmohammad$^{1,}$\footnote{${}^{*}$ Authors with equal contributions. } , 
Torsten Held$^{2,*}$, 
Michael L\"assig$^{2}$}

\date{\small $^1$ Joseph-Henri Laboratories of Physics and Lewis-Sigler Institute for Integrative Genomics, \\ Princeton University, Princeton, NJ 08544, USA\\ $^2$ Institut f\"ur Theoretische Physik, Universit\"at zu K\"oln, \\ Z\"ulpicherstr. 77, 
50937, K\"oln, Germany
}

\maketitle

\begin{abstract} 
Molecular traits, such as gene expression levels or protein binding affinities, are increasingly accessible to quantitative measurement by modern high-throughput techniques. Such traits measure molecular functions and, from an evolutionary point of view, are important as targets of natural selection. We review recent developments in evolutionary theory and experiments that are expected to become building blocks of a quantitative genetics of molecular traits. We focus on {\em universal} evolutionary characteristics: these are largely independent of a trait's genetic basis, which is often at least partially unknown. We show that universal measurements can be used to infer selection on a quantitative trait, which determines its evolutionary mode of conservation or adaptation. Furthermore, universality is closely linked to predictability of trait evolution across lineages. We argue that universal trait statistics extends over a range of cellular scales and opens new avenues of quantitative evolutionary systems biology.
\end{abstract}
\end{titlepage}

\section*{Introduction}

Quantitative traits are important links between genotypes, organismic functions, and fitness. For some molecular traits, recent sequence data and high-throughput trait measurements have produced quantitative genotype-phenotype maps. Examples include the sequence-dependent binding of transcription factors and histones to DNA, and the formation of RNA secondary structures. For the vast majority of complex traits, however, quantitative genotype-phenotype maps are out of reach. Even comparatively simple molecular traits, such as gene expression levels, depend on a mosaic of cis- and trans-acting sequence loci. We do not know their precise numbers, positions and trait amplitudes, nor relevant evolutionary rates such as the amount of recombination between these loci~\cite{Rockman11}. This lack of knowledge begs an obvious question: Which evolutionary properties of a quantitative trait are {\em universal}, that is, independent of these molecular details? In particular, can we formulate natural selection on quantitative traits and their resulting modes of evolution independently of their genetic basis? This article is on universality in molecular evolution. We introduce universality as an emerging statistical property of complex traits, which are encoded by multiple genomic loci. We give examples of experimentally observable universal trait characteristics, and we argue that universality is a key concept for a new quantitative genetics of molecular traits. Three aspects of this concept are discussed in detail. First, universal statistics governs evolutionary modes of conservation and adaptation for quantitative traits, which can be used to infer natural selection that determines these modes. Furthermore, there is a close link between between universality and predictability of evolutionary processes. Finally, universality extends to the evolution of higher-level units such as metabolic and regulatory networks, which provides a link between quantitive genetics and systems biology.

\section*{Universality in molecular evolution}

In a broad sense, universality means that properties of a large system can become independent of details of its constituent parts. This term has been coined in statistical physics, where it refers to macroscopic properties of large systems that are independent of details at the molecular scale~\cite{Kadanoff1990}. For example, the amount of fluid running through a tube per unit time depends only on the viscosity of the fluid, the diameter of the tube, and the pressure gradient, but not on the detailed chemical composition of the fluid. Thus, rather different fluids have the same flux properties as long as their viscosity is the same. This a strong, experimentally testable statement. It is not always true: if the tube narrows  at some point into a nozzle, the fluid becomes turbulent and other things besides viscosity matter. This tells another upfront message: universality is usually not a mathematical identity, but an approximation that is accurate in some cases but not in others. 

Universality also arises in evolutionary biology. As in physics, it is a property of systems with a large number of components, and it has strong consequences for experiment and data analysis. In the following, we will discuss a number of examples, and we will pinpoint these components and experimental consequences in each case. 

In population genetics, Kimura's celebrated diffusion model for the evolution of allele frequencies is a universal description~\cite{Kimura:1964}. The Kimura model predicts that the frequency distribution of mutant alleles in a large population depends only on the size of the population and the selection coefficients of the alleles, but not on the details of the reproductive process of individuals. Many more detailed models of reproduction, including the Wright-Fisher process~\cite{Moehle2001}, the Moran process~\cite{Moran1985}, and branching processes~\cite{KarlinMcGregor1964}, have a common diffusion limit in large populations. Importantly, the universal frequency spectra of the Kimura model are statistical quantities; observing such spectra requires frequency data from a large number of segregating alleles in a population. Hence, the universal spectrum most frequently observed in genomic data is the famous inverse-frequency form for synonymous alleles, which evolve near neutrality. For alleles under selection, universality is often confounded by the heterogeneity of selection coefficients at different genomic loci.

Universality may arise even in Darwinian evolution under strong selection, for example, in rapidly adapting asexual populations. Due to the lack of recombination,  competition between simultaneously  spreading beneficial mutations leads to complex patterns of rise and fall in their population frequencies. However, if an adaptive process is carried by a large number of  segregating  alleles, it can be described in a simpler way by a so-called traveling fitness wave~\cite{Tsimring,Rouzine:2003en,Wilke:2004gs,Desai:2007wv,Rouzine2008,Fisher2011,Hallatschek2011,Good2011,Lassig:2012gi,Fisher2013}. The speed of this wave, which is also referred to as fitness flux~\cite{Mustonen:2010ig},  becomes a universal quantity: it depends primarily on rate and average effect of beneficial mutations, but not on the detailed distribution of their selection coefficients~\cite{Desai:2007wv,Fisher2011,Good2011,Fisher2013}. In other words, the fitness flux decouples from details of the underlying genomic evolution. More generally, the distribution of fixed mutations  becomes insensitive to the details of genomic fitness effects  and can be characterized by only a few effective parameters~\cite{Desai:2007wv, Good2011,Fisher2013}. This feature has also  been observed  in a microbial evolution  experiment under strong selection pressure~\cite{Hegreness:2006ej}. Another striking universal feature emerges for  ``passenger''  mutations carried to fixation by hitchhiking with linked beneficial alleles:  their substitution rate becomes independent of their selection coefficients and close to the neutral mutation rate~\cite{Schiffels:2011fua, Good2011, Lassig:2012gi}. This effect increases the substitution rate of deleterious mutations, which may have significant impact on the adaptive dynamics of pathogens~\cite{Strelkowa12} and on cancer progression~\cite{McFarland13}. 
Universality has an important consequence for theory: it may allow the construction of models that are simple enough to be solvable, but share their universal properties with more realistic models. In this way, adaptive evolution of asexual populations has recently been mapped onto a solvable stochastic traveling-wave model~\cite{Hallatschek2011}. As in physics, this universality has its limitations. For example, if just a few beneficial alleles coexist at a given point in time, the fitness wave starts to stutter and its speed changes~\cite{Schiffels:2011fua}. 

Averaging of allelic contributions is a generic feature in the evolution of quantitative traits. For complex traits, which are encoded by a sufficiently large number of genomic loci, this results in universality. Consider, for example, R.A. Fisher's classic geometric model, which describes the evolution of a trait with $d$ components in a single-peak quadratic landscape of (log, i.e., Malthusian) fitness~\cite{Fisher:1930wy}. In this model, selection favors a unique optimal trait value, but deleterious mutations, which can fix by genetic drift, cause the trait to scatter at some distance from the optimum. This process reaches a selection-mutation-drift equilibrium that depends on the number of trait components, the effective population size, the mutation rate, and the strength of stabilizing selection, but not on details of the genomic loci and their evolution. In this case, the reason underlying universality is compensatory evolution caused by stabilizing selection: individual loci behave in a highly stochastic way, but deleterious changes at one locus tend to be offset by simultaneous or subsequent beneficial changes at other loci. Perhaps the simplest measurable universal quantity is the expected fitness cost or {\em genetic load} $\L \simeq d/4N$, where $N$ denotes the effective population size. This formula characterizes evolutionary equilibrium for low mutation rates and sufficiently strong stabilizing selection. We derive it in Box~1; variants have been obtained previously in refs.~\cite{Hartl1998,Poon:2000vg,Tenaillon:2007jda}. 

A number of recent studies have treated quantitative traits under stabilizing selection, emphasizing the ``coarse-graining" from genomic alleles to trait variables and the analogy to statistical mechanics ~\cite{Barton:1986,Berg:2004dz,ML05,Lassig:2007iq,Barton:2009di,Barton:2009genetics,Neher:2011wc,deVladar:2011bs}. Other selection scenarios that have been explored include directional selection~\cite{Barton:2009genetics}, adaptation to a moving fitness optimum~\cite{KoppHermisson,deVladar:2011bs} and apparent selection in the presence of pleitropy \cite{McGuigan:2011fk}. Related statistical methods for the analysis of complex traits are of growing interest for genome-wide association studies~\cite{GenomesProjectConsortium:2010gj,Rockman11,Tennessen:2012ck,Kiezun:2012dw}. 
\begin{figure}[t]
\fbox{
 \parbox{0.97\textwidth}{
\small 
\subsection*{\bf Box 1: Genetic load is universal}
  
Genetic load, which is defined as the difference between the maximum (Malthusian) fitness
and the mean fitness in a population, 
\[
 \L =  f^* - f_{\rm mean}. 
\]
At evolutionary equilibrium in a fitness landscape, the dominant contribution for low mutation rates is the drift load, which is generated by deleterious substitutions at the trait's constitutive sites. The average drift load is easy to compute from the equilibrium distribution of the trait mean, which takes the ``Boltzmann'' form 
\[
Q(\Gamma) = \exp[2 N f (\Gamma) + S(\Gamma)], 
\]
where the {\em entropy} $S(E)$ is the log density of sequence states mapping onto a given trait value~\cite{Lassig:2007iq,Nourmohammad:2013ty}. Defining a trait-dependent {\em free fitness} $\tilde f(E) = f(E) + (1/2N) S(E)$, we can write the equilibrium distribution in the form $Q(\Gamma) = \exp[2 N \tilde f (\Gamma)]$; the underlying formalism in sequence space has been developed in refs.~\cite{Iwasa:1988ws, Berg:2004dz, Sella:2005da, Mustonen:2010ig}. Given a quadratic form of $f(E)$ and of $S(E)$ (which often is a good approximation), we obtain an explicit expression for the load in free fitness,
\[
2N \big (\max_\Gamma \tilde f (\Gamma) - \langle \tilde f \rangle \big ) = \frac{1}{2}, 
\]
which reduces to $\langle 2 N \L \rangle \simeq 1/2$ in the strong-selection regime $c_0 \equiv f'' (\Gamma^*) \gg | S'' (\Gamma^*) | /2N$~\cite{Nourmohammad:2013ty}. Drift load, in particular,  does not depend on the number of trait loci, $\ell$ (in contrast to mutational load, which increases with~$\ell$). For a $d$-component trait as in  Fisher's geometric model, this formula generalizes to the form quoted in the text, $\langle 2 N \L \rangle \simeq d/2$. It is a direct evolutionary analogue of the equipartition theorem in statistical thermodynamics, which states that every degree of freedom that enters the energy function quadratically contributes an average of $k_B T/2$ to the total energy of a system at temperature $T$ (the proportionality factor $k_B$ is Boltzmann's constant).

Generic evolutionary processes of quantitative traits have two additional load components, which may become dominant over drift load: the diversity load for polymorphic traits, which is  proportional to $\mu$, and the adaptive load in a fitness seascape, which arises from the lag of the population behind the moving fitness peak  and is proportional to  $\v/\mu$ (where $\v$ is the driving rate defined in the text)~\cite{Held13}. 
The different scaling of these load components with the mutation rate $\mu$ expresses a generic feature of adaptive processes: higher mutation rates increase the equilibrium load, but facilitate adaptive changes. 
}}
\end{figure}

All of these studies cover specific classes of quantitative traits, which is reflected in their assumptions on genome evolution. One group of models applies to {\em microscopic} quantitative traits, which depend only on a few genomic sites and are generically monomorphic in a population~\cite{Berg:2004dz, ML05, Lassig:2007iq, Kinney:2008tb}. For such traits, a population can be approximated as a point in trait space that moves by beneficial and deleterious substitutions; trait diversity and linkage disequilibrium between trait loci are negligible. An example of a microscopic trait is the sequence-dependent binding free energy of a transcription factor to its DNA target sites~\cite{BH}. Other models treat a complementary class of {\em macroscopic} or {\em polygenic} traits, which are encoded by many genomic loci and are always polymorphic. In the spirit of classical quantitative genetics, these models assume fast recombination between the trait loci, which results in complete linkage equilibrium~\cite{Lynch:1998vx,Barton:1986,Barton:2009di,Barton:2009genetics,deVladar:2011bs}. Interestingly, the evolutionary statistics of a polygenic trait under stabilizing selection can be derived from a maximum-entropy principle, and some of this statistics extends beyond evolutionary equilibrium to the case of a moving fitness peak~ \cite{deVladar:2011bs}. Small deviations from linkage equilibrium due to insufficient recombination can be treated by a perturbative approach~\cite{Neher:2011wc}. However, lowering the recombination rate can have more drastic effects: at some point, recombination ``freezes'' and the reproductive process becomes essentially asexual~\cite{Neher2009,  Neher:2011wc,Neher:2013dj}.

A more comprehensive quantitative genetics of molecular traits requires tractable models that cover the many important molecular phenotypes outside the above classes. Such {\em mesoscopic} traits are encoded by multiple genomic sites and are generically polymorphic, yet these sites are located in a confined genomic region and evolve under at least partial genetic linkage~\cite{Comeron:2002vn}. Mesoscopic traits are often building blocks of large-scale organismic traits. Examples are gene expression levels, which depend on multiple cis- and trans-acting sites~\cite{Ptashne}, protein and RNA fold stabilities, which depend on the coding sequence of a single gene~\cite{Smith:1970uz,Fernandez:2011uj} and histone-DNA binding energies, which are determined by segments of about 150 contiguous base pairs~\cite{Radman:2010, Weghorn:2013cy}. Mesoscopic traits raise another universality question: are there evolutionary features of a trait that depend only weakly on the amount of recombination between its constitutive sites? This question is addressed in a recent study, which establishes an analytic evolutionary theory of quantitative traits in asexual populations and identifies key trait observables that are largely independent of the recombination rate over a wide range of evolutionary parameters~\cite{Nourmohammad:2013ty}. Yet another dimension of universality comes into play if we look at adaptive evolution of quantitative traits. To describe the selective cause of long-term adaptive processes, we must generalize static fitness landscapes to fitness {\em seascapes} that change over macro-evolutionary periods~\cite{Lassig:2007vb,Mustonen08,ML09}. The simplest such seascape is a moving fitness peak, which drives adaptive trait changes towards a time-dependent optimal value. Are there features of the adaptive process that depend only on broad seascape characteristics such as the mean squared peak displacement, but are independent of size and timing of individual peak changes? This question is addressed in another recent study, where we develop a non-equilibrium theory of quantitative trait evolution~\cite{Held13}. 

Together, these developments show that universality characterizes not just individual features of a trait, but its entire evolutionary mode of conservation or adaptation. To develop this picture, we will now introduce minimal land- and seascape models that describe selection on a quantitative trait in a universal way.

\section*{Evolutionary modes of quantitative traits}

Complex molecular traits, such as the examples of the previous section, are encoded by multiple genomic sites and are generically polymorphic. The following discussion will focus on additive complex traits, for which the trait value of an individual is the sum of the allelic contributions at the trait's constitutive sites. Clearly, this additivity assumption does not exclude fitness interactions (epistasis) between these sites; as we will argue below, such interactions are indeed a generic feature of quantitative traits. The distribution of trait values in a population often is approximately Gaussian. It is then well characterized by just two numbers: its mean, $\Gamma$, and its total heritable variance, $\Delta$, which will be referred to as the trait diversity. Mean and diversity evolve by selection on the trait, by mutations and genetic drift at its constitutive sites, and, in a sexually reproducing population, by recombination between these sites.  The change of the trait distribution with evolutionary time $t$ defines the trait divergence  $D(\tau) = (\Gamma(t+\tau) - \Gamma (t))^2$ as the squared displacement of the trait mean over a macro-evolutionary period $\tau$. Because this dynamics is stochastic, evolutionary theory always addresses an ensemble of evolving populations; individual populations differ in their realizations of the stochastic evolutionary processes. Expectation values in this population ensemble, such as the average divergence $\langle D(\tau) \rangle$ and the average diversity $\langle \Delta \rangle$, should be compared with the corresponding data averages over parallel evolving populations.  

To define evolutionary modes of quantitative traits, we relate their statistics of divergence and diversity to the underlying natural selection. We use a minimal model chosen for its conceptual and computational simplicity: fitness depends quadratically on the trait coordinate $E$, and the trait value of maximum fitness, $E^*$, is a function of evolutionary time~$t$,  
\EQ
f(E,t) = f^* - c_0 \big (E - E^*(t) \big )^2. 
\label{fitness}
\EE
Despite its simplicity, this fitness function covers a broad spectrum of evolutionary scenarios. For constant trait optimum $E^*$, it is a time-honored model of {\em stabilizing selection}, which is already included in Fisher's geometric model~\cite{Fisher:1930wy,Lynch:1998vx}. Nearly all known examples of empirical fitness landscapes for molecular quantitative traits are of  single-peak~\cite{Poelwijk:2011ce}  or mesa-shaped~\cite{Gerland03,ML05, Kinney:2008tb,Wylie:2011io,Hermsen:2012fz} forms. Mesa landscapes describe directional selection with diminishing return: 
they contain a fitness flank on one side of a characteristic ``rim'' value $E^*$ and flatten to a plateau of maximal fitness on the other side. Furthermore, trait values on the fitness plateau tend to be encoded by far fewer genotypes than low-fitness values. This differential coverage of the genotype-phenotype map turns out to generate an effective second flank of the fitness landscape, which makes our subsequent theory applicable to mesa landscapes as well\footnote{
Mathematically, we can define a phenotype-dependent {\em free fitness} $\tilde f(E)$, as discussed in Box~1. This function turns out to be again of the form (\ref{fitness}), with a maximum close to the fitness rim $E^*$. }. For time-dependent $E^*(t)$, equation (\ref{fitness}) becomes a fitness seascape model with a component of {\em directional selection}  that is generated by displacements of the fitness peak in the trait coordinate\footnote{
In this seascape model, directional selection acts only on the trait mean $\Gamma$. Because $c_0$ is a constant, the diversity $\Delta$ remains in evolutionary equilibrium under stabilizing selection, and its expectation value $\langle \Delta \rangle$ is time-independent. 
}. 
These displacements can be diffusive, reflecting continuous incremental ecological changes affecting the trait. They can also include rare large-amplitude shifts, which are caused by major ecological events such as migrations or speciations. 

A key feature of our fitness model is that it provides universal selection parameters. First, how strong is stabilizing selection? This should clearly be characterized by a universal number, which does not depend on individual trait loci. In equation (\ref{fitness}), however, the strength of selection remains undefined as long as the scale used to measure trait values is arbitrary; for example, we can use centimeters or inches to measure body height. A natural trait scale $E_0$ can be defined, for example, as the width of the total trait repertoire under neutral evolution\footnote{
More precisely, this scale is given by the trait asymptotic divergence, $E_0^2 = (1/2) \lim_{\tau \to \infty} D_0 (\tau)$, in the regime of low mutation rate, $\mu N \ll 1$ (the subscript $0$ refers to neutral evolution, $c_0 = 0$). For additive traits, there is an equivalent micro-evolutionary definition in terms of the average trait diversity under neutral evolution, $E_0^2 = \langle \Delta \rangle_0 / (4 \theta)$, where $\theta = \mu N $ denotes the neutral nucleotide diversity.
}.
Measuring trait values in units of $E_0$ and fitness per $(2N)$ generations, we obtain a scaled selection constant $c = 2N E_0^2 c_0$, which is a dimensionless measure of stabilizing strength. This constant equals the ratio of the neutral trait variance $E_0^2$ and the weakly deleterious trait variance in the fitness landscape $f(E)$, which, by definition, produces a fitness drop $\leq 1/2N$ below the maximum $f^*$. In other words, stabilizing selection is substantial if $c \gtrsim 1$~\cite{Nourmohammad:2013ty}. This selection regime defines the evolutionary mode of {\em trait conservation}, which is illustrated in Fig.~1(a). At the genomic level, the evolutionary dynamics involves negative epistasis and compensatory allele changes at the constitutive sites, which are an immediate consequence of the fitness nonlinearity in equation~(\ref{fitness}). This dynamics generates substantial constraint on the expected trait divergence $\langle D(\tau) \rangle$ over macro-evolutionary periods, as we will discuss in the next section.  

How important are fitness peak displacements? If we are not interested in details of individual peak shifts, we can promote the model~(\ref{fitness}) to a stochastic fitness seascape, which is characterized a single additional parameter $\v$.  This parameter is defined as the mean squared peak displacement, measured in units of $E_0^2$ and per unit of evolutionary time, and is called the {\em driving rate} of selection~\cite{Held13}. Our population ensemble now includes the stochasticity of selection; that is, individual populations of this ensemble differ in the realizations of the peak displacement process. If this process is Markovian, the mean squared peak displacement\footnote{
This linear form applies in a short-time regime; in addition, it is convenient to introduce a saturation term in the long-time limit.
} 
over a period $\tau$ is simply $E_0^2 \v \tau$. The regime of substantial and time-dependent selection ($c \gtrsim 1$, $\v > 0$) defines the evolutionary mode of {\em trait adaptation}, which is illustrated in Fig.~1(b). Here we limit the discussion to {\em macro-evolutionary} fitness seascapes, which have low enough driving rates to allow for efficient adaption of the trait mean.
The mathematical condition for this selection regime is quite intuitive: the mean square peak displacement per coalescence time of $(2N)$ generations has to be smaller than the average trait diversity, $\v \lesssim \tilde \v \equiv \langle \Delta \rangle / (2N E_0^2)$. 
As detailed in the next section, adaptation in a macro-evolutionary fitness seascape leads to an expected trait divergence $\langle D(\tau) \rangle$ with constraint on shorter scales and adaptive increase on longer scales of evolutionary time~\cite{Held13}. This behavior is universal in two ways: it depends neither on genetic details of the trait nor on dynamical details of the fitness seascape. Whether adaptation plays a substantial role turns out to depend on the fitness parameters $c$ and $\v$ and on the evolutionary period $\tau$; this point will be discussed further below. 

In summary, the fitness seascape model (\ref{fitness}) provides a unifying framework for the analysis of selection on quantitative traits. Importantly, it shows that stabilizing and directional selection are not mutually exclusive, but joint features of dynamic selection. The model has two universal parameters, the stabilizing strength $c$ and the driving rate $\v$, which define evolutionary modes of quantitative traits. Due to the simplicity of the model, its selection parameters and the resulting evolutionary modes are expected to be meaningful measures for a large number of quantitative traits. We now show how these modes can be inferred using universal relationships between fitness seascapes and trait data.

\begin{figure}[t]
\boxed{\includegraphics[width=\textwidth]{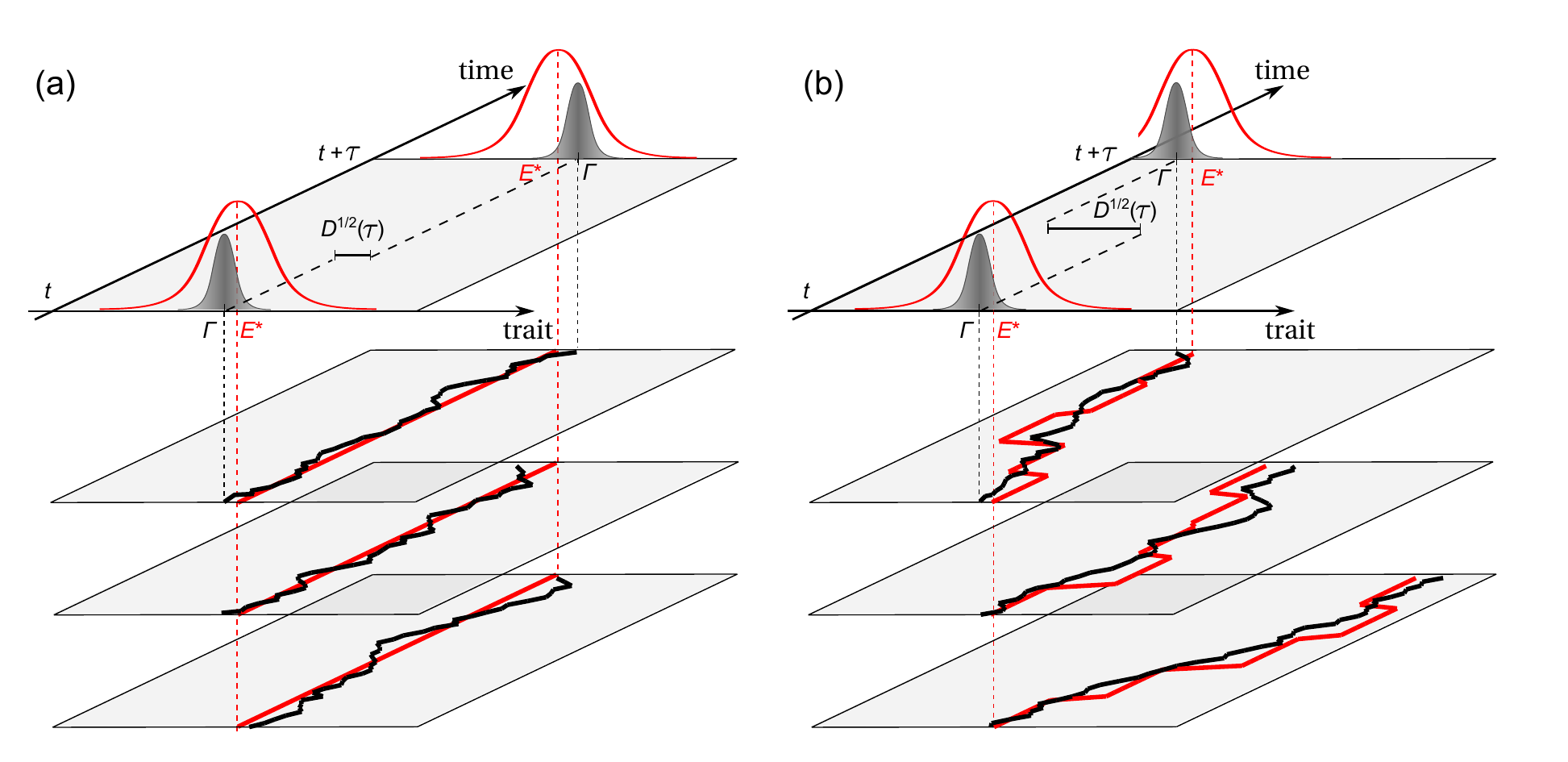}}

\caption{{\bf Conservation and adaptation of quantitative traits (schematic).} 
(a) {\em Evolution under stabilizing selection. }
Upper panel: The distribution of trait values $E$ in a population (gray filled curves) evolves in a fitness landscape $f(E)$ (thin red curves) with a time-independent optimum trait value  $E^*$. The trait divergence $D (\tau) = (\Gamma (t + \tau) - \Gamma (t))^2$ over a macro-evolutionary period $\tau$ results from reproductive fluctuations (genetic drift) of the trait mean $\Gamma$ around the optimum $E^*$. 
Lower three planes: The theory of this process describes an ensemble of populations; the evolution of the trait mean (black curves) around the fixed optimum (red lines) is shown for three individual populations from this ensemble. These populations differ in their realizations of genetic drift. 
(b)~{\em Adaptive evolution. }
Upper panel: The trait distribution evolves in a fitness seascape $f(E,t)$ with a time-dependent optimum value $E^*(t)$. The trait divergence $D(\tau)$ results from adaptive changes of the trait mean $\Gamma$, which follow displacements of the fitness peak $E^*(t)$, as well as from genetic drift of $\Gamma$.
Lower three planes: In a stochastic fitness seascape, individual populations from the ensemble differ in  their realizations of peak displacements (red curves) and of genetic drift. }
\end{figure}

\section*{Inference of conservation and adaptation}

Most selection inference methods use genomic information. The well-known McDonald-Kreitman test, for example, is based on differences in diversity and divergence statistics between non-synonymous mutations and the synonymous ones which are assumed to evolve near neutrality~\cite{Kreitman:1991vh}. The situation is different for quantitative traits: in general, their constitutive genomic sites are at least in part unknown, and we do not have a corresponding ``null trait'' that evolves near neutrality. As a consequence, the neutral trait scale $E_0$ used in our definition of evolutionary modes is unknown as well. But the absence of a neutral gauge is more fundamental: for many quantitative traits, neutral evolution is not only practically, but also conceptually inadequate as a null model. For example, a neutrally evolving gene expression level is zero, because genes are rapidly converted to pseudogenes in the absence of selection. Difficulties of this kind are in part responsible for controversial results on selection and adaptation of gene expression levels~\cite{Khaitovich:2006bi,Gilad:2006kt,Bedford:2009fy,Fraser:2010gm}.

To infer selection on quantitative traits without reference to a neutral gauge, we use the fitness seascapes of equation (\ref{fitness}) as a family of minimal selection models. These models depend on two parameters $c$ and $\v$, which can be calibrated with trait divergence and diversity data. Specifically, our inference method is based on the ratio between divergence and diversity,  
\EQ
\Omega (\tau)\equiv2\theta \frac{\langle D (\tau) \rangle}{ \langle \Delta \rangle}. 
\label{omega}
\EE   
This ratio is normalized by multiplication with twice the nucleotide diversity $\theta = \mu N$.  It turns out to be a universal function of divergence time, which does not depend on the trait scale $E_0$~\cite{Held13}. As shown in Fig.~2, this function depends on selection in two characteristic ways: first, stabilizing selection always affects trait divergence more strongly than diversity; second, the trait divergence in a macro-evolutionary seascape has a short-time regime dominated by genetic drift and a long-time regime dominated by adaptation. The quantitative behavior of $\Omega (\tau)$, which is detailed in Box~2, can serve to infer the selection parameters $c$ and $\v$, and the resulting mode of trait evolution. The inference uses divergence and diversity data that are averages over parallel measurements in lineages of diverging species or colonies. This is a typical kind of measurement, for example, in microbial evolution experiments~\cite{ Barrick:2009in,Lang:2013fha}. The underlying analytical theory, which has been presented here for evolution over a single period $\tau$, can be generalized to an entire phylogeny. It can also be generalized to further seascape classes and evolutionary modes, for example adaptive evolution of the trait diversity in response to time-dependent stabilizing strength $c(t)$. 

\begin{figure}[t]
\boxed{
\includegraphics[width=.5\textwidth]{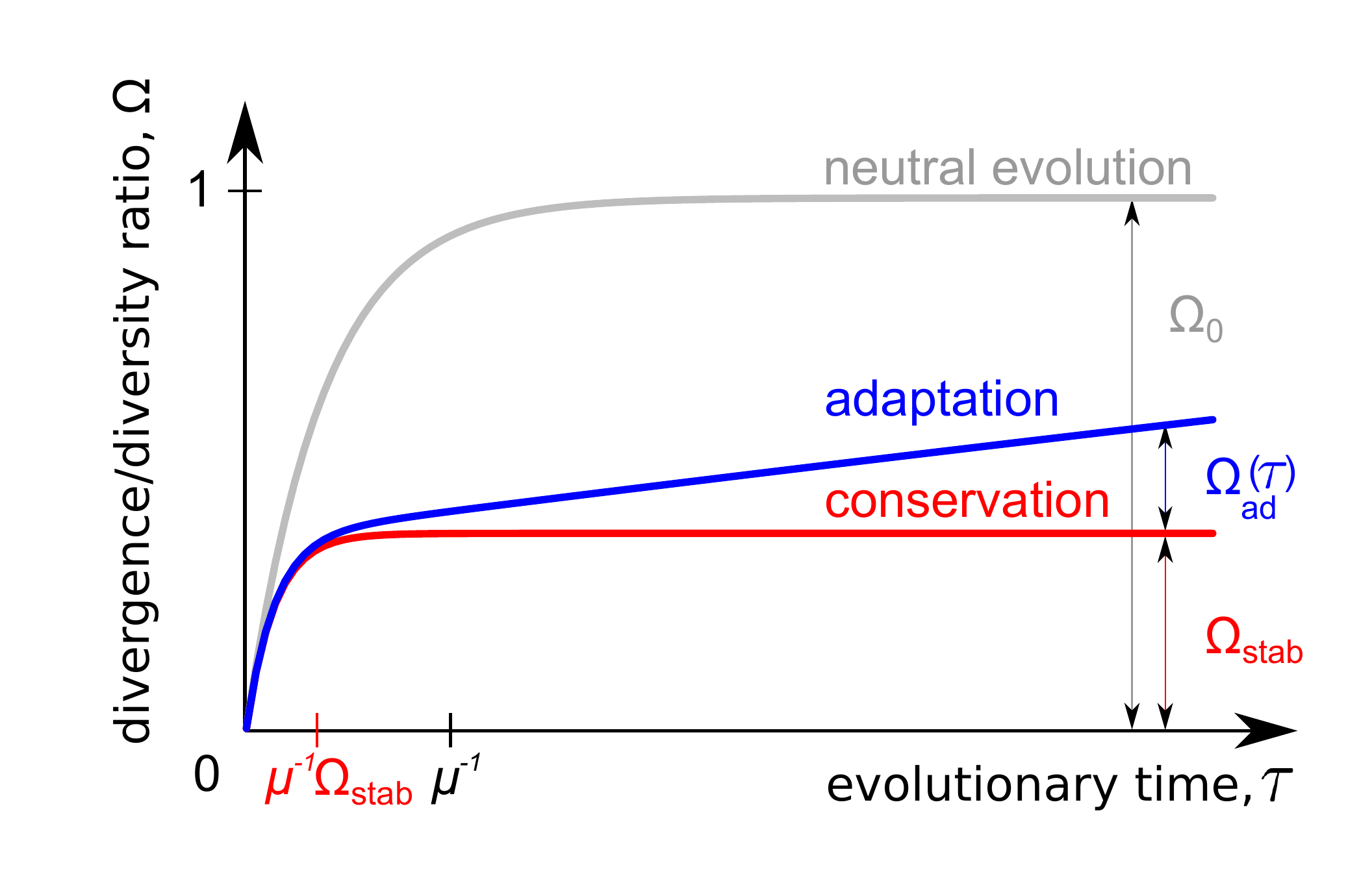}
}
\caption{{\bf Evolutionary modes and inference of selection for quantitative traits.}  The figure shows the universal divergence-diversity ratio $\Omega$, as defined in eq.~(\ref{omega}), for a quantitative trait evolving in a single-peak fitness land- or seascape. This ratio is plotted as a function of evolutionary time $\tau$. {\em Neutral evolution}: The function $\Omega (\tau)$ reaches the saturation value $\Omega_0 = 1$ for large times (grey curve). {\em Conservation}: This function has a smaller saturation value $\Omega_\stab$, which is reached faster than for neutral evolution (red curve). {\em Adaptation}: There is a linear surplus $\Omega_\ad (\tau)$, which measures the amount of adaptation (blue curve). This behavior can be used to infer selection, as detailed in Box~2.  }
\end{figure}

\begin{figure}[t]
\fbox{
 \parbox{0.97\textwidth}{
\small 
\subsection*{\bf Box 2: Universal statistics of trait divergence and diversity}
        
The scaled divergence-diversity ratio $\Omega (\tau)$ of a quantitative trait, as defined in eq.~(\ref{omega}), is a universal function of evolutionary time. Its form reflects the mode of trait evolution. In a single-peak fitness seascape with stabilizing strength $c$ and driving rate $\v$, we can distinguish the following modes~\cite{Nourmohammad:2013ty, Held13}:
\begin{itemize}
\item{\em Neutral evolution} ($c = 0$): The function $\Omega (\tau | c \!=\! 0, \v  \!=\! 0)$ has initial slope 1 and increases up to an asymptotic value $\Omega_0 = 1$ with a saturation time $1/\mu$. 

\item{\em Conservation in a fitness landscape}   ($c \gtrsim 1$, $\v = 0$): Under substantial stabilizing selection, the function $\Omega (\tau | c , \v  \!=\! 0)$ has the same initial slope 1 and increases up to an asymptotic value $\Omega_\stab < 1$ with a proportionally shorter saturation time $\Omega_\stab / \mu$. There are two relevant selection regimes: $\Omega_\stab \approx 1/(2c)$ for intermediate stabilizing strength ($1 \lesssim c \lesssim 1/ \theta$), and $\Omega_\stab \approx \sqrt{\theta/c}$ for asexual evolution under extremely strong selection ($c  \gtrsim 1/ \theta$); universality with respect to recombination rate is broken in the strong selection regime.  

\item{\em Adaptation in a macro-evolutionary fitness seascape}  ($c \gtrsim 1$, $0 < \v \lesssim \tilde \v$): The divergence-diversity ratio takes the form $\Omega (\tau | c, \v) = \Omega(\tau | c, \v \!= \! 0) + \Omega_\ad (\tau | \v)$, with
$\Omega_\ad (\tau | \v) = (\v / \tilde \v) (\mu \tau)\, [1 + O(\v \tau)]$. This form displays two macro-evolutionary regimes: the drift-dominated regime for $\tau \lesssim \Omega_\stab / \mu$ and the adaptive regime for $\tau \gtrsim \Omega_\stab / \mu$. In the latter regime, the ratio of the two $\Omega$ components is related to the cumulative fitness flux~\cite{Held13},
\[
\frac{\Omega_\ad (\tau)}{\Omega_\stab} \approx \v c \tau  = \langle 2 N \Phi (\tau) \rangle.  
\]
\end{itemize}
}}
\end{figure}

How much adaptive trait evolution has happened over a macro-evolutionary divergence period? The most natural measure of adaptation is the {\em fitness flux} $\phi (t)$, which is defined as the rate of  movement on a fitness land- or seascape by genotype or phenotype changes in a population~\cite{Lassig:2007vb,Mustonen:2010ig}. The cumulative fitness flux, $\Phi(\tau) = \int_{t}^{t + \tau} \phi (t) \, dt$, measures the total amount of adaptation over a macro-evolutionary period in a population history. This quantity satisfies the fitness flux theorem, which generalizes Fisher's fundamental theorem of natural selection to generic mutation-selection-drift processes~\cite{Mustonen:2010ig}. As shown by the fitness flux theorem, the average cumulative fitness flux over parallel evolutionary histories, in units of $1/2N$, measures the importance of adaptation compared to genetic drift: adaptation is substantial if $\langle 2 N \Phi (\tau) \rangle \gtrsim~1$. For the evolution of a quantitative trait in a fitness seascape of the form (\ref{fitness}), we can write the fitness flux to a good approximation as the evolutionary rate of the trait mean multiplied by the local seascape gradient,
\EQ
\phi (t) =\frac{d \Gamma (t)}{d t} \times  \frac{\partial f(\Gamma, t)}{\partial \Gamma}. 
\EE
In the weak-mutation regime, the average cumulative fitness flux takes the simple form $\langle 2 N \Phi (\tau) \rangle \approx \v c \tau $~\cite{Held13}. This quantity depends only on the stabilizing strength, $c$, and the scaled mean squared displacement of the fitness peak, $\v \tau$, but not on details of the peak dynamics. It is universal and, in accordance with the fitness flux theorem, always non-negative~\cite{Mustonen:2010ig}. As shown in Box~2, the scaled flux  $\langle 2 N \Phi (\tau) \rangle$ is related to the divergence data of Fig.~2: it approximately equals the ratio $ \Omega_\ad(\tau) / \Omega_{\stab}$.

\section*{Predictability of genotypic and phenotypic evolution}

Chance and necessity are venerable topics in evolutionary biology. Modern evolution experiments with microbial and viral systems can address these topics from a new angle, because the same experiment can be run for many populations in parallel~\cite{Lenski:1994vta}. We can then ask how repeatable the experimental outcome is. In other words, can the evolution of a population  be predicted from the knowledge of a previous experiment? This question can be addressed at the level of genotypes, of phenotypes, and of fitness. At each level, we can assess the predictability of the final adaptive outcome or of the complete mutational trajectory over the course of evolution.

Given the combinatorial complexity of genotype space, genotypic predictability is low unless we restrict our focus to small subspaces, such as the adaptive genotypes of a single protein that acquires antibiotic resistance~\cite{Weinreich:2006ig}. But even in such restricted spaces, the statistics of adaptive processes and their mutational trajectories is complex~\cite{Szendro:2013uj}. Because genotype-dependent fitness landscapes are often rough, the rate and paths of adaptation depend on two opposing forces: higher supply of single beneficial mutations canalizes the dynamics along accessible paths and increases predictability, while higher supply of double mutations bridges intermediate fitness valleys and decreases predictability. As shown by a recent computational study, these dynamics can generate a non-monotonic dependence on experimental control parameters such as population size: the predictability of final genotype is generically low, but there are sweet spots of high predictability at specific values of these parameters~\cite{Szendro13}.  

At the genome-wide level, a similarly complex pattern of predictability emerges for beneficial mutations at individual genomic sites. A recent experimental study of adaptive evolution in 40 yeast populations identified two opposing forces influencing the substitution rates of beneficial mutations: their origination rate increases, but the probability of surviving clonal interference decreases with increasing population size~\cite{Lang:2013fha}. This is in accordance with clonal interference theory: the substitution rate of mutations with a given selection coefficient is predicted to depend on population size in a non-monotonic way, which again leads to a sweet spot of predictability for some value of $N$~\cite{Schiffels:2011fua}. This non-monotonic behavior signals drastic deviations of the evolutionary dynamics from the classical picture of rare, independent beneficial mutations (which have substitution rates proportional to $N$ and to their selection coefficient). 

Parallel and convergent evolution has been identified in several studies at the level of genes and higher genomic units; this is often coupled with strongly divergent evolution of individual mutations in these genes~
\cite{Cooper03, Blount08,Barrick:2009in, Saxer2010,  Wichman10}.
Do the mutations in a given gene affect a common trait, say, the gene's enzymatic or regulatory activity? This point has been highlighted in some recent studies. A massively parallel evolution experiment, which involved 115 lines of {\em E. coli} adapting to high temperature, revealed strikingly convergent evolution of molecular traits, including complex traits such as RNA polymerase function~\cite{Tenaillon2012}. Another parallel experiment explored resistance of {\em E. coli} to three  different types  of antibiotics~\cite{Toprak:2012ff}. Parallel populations had similar phenotypic trajectories during their adaptive evolution  in response to  all three types of  antibiotics. Genotypic parallelism was found in response to one of the drugs, suggesting a more constrained mutational pathway in that case. Remarkably, high phenotypic predictability is not limited to the specific types of selection imposed in laboratory experiments. A recent study reported similarly high  predictability of specific adaptive traits for {\em E.~coli} populations in the highly complex ecosystem of the mouse gut~\cite{Barroso13}. 

We now show that high phenotypic predictability, coupled with low genotypic predictability, is a generic, universal feature in the evolution of complex quantitative traits. Recall the simple reason underlying universality: Stabilizing selection generates compensatory genetic changes at the trait's constitutive loci. That is, stochastic changes in trait and fitness at one locus tend to be buffered by simultaneous and subsequent changes at other loci. This effect increases with the complexity of the quantitative trait, i.e., with the number of sequence loci it depends upon. Stabilizing selection can generate substantial evolutionary constraint of trait mean and diversity, even if selection on individual loci is weak. Thus, the phenotypic constraint does not depend on details of any single locus, but is an emerging property of all of the trait's constitutive loci. To estimate the predictability of a trait evolving in a fitness seascape of the form (\ref{fitness}), consider the schematic evolution experiment sketched in Fig.~1. We look at an ensemble of populations that start from a common trait distribution and evolve in a given fitness landscape or in individual realizations of a fitness seascape. We can ask to what extent the trait repertoire or the underlying genotype repertoire of one evolved population is predictive of the trait or genotype values in another evolved population, depending on the evolution period $\tau$. For complex traits, it is clear that the predictability of a trait's constitutive genotypes is  generically poor, because many different genotypes lead to a trait value close to a given fitness optimum. This degeneracy is a major limiting factor for gene association studies. At the phenotypic level, we can estimate predictability by comparing the expected trait diversity in a single population with the mean squared trait distance between two evolved populations  \cite{Nourmohammad:2013ty},
\EQ
\mathcal P (\tau) =  \sqrt{\frac{ \langle \Delta \rangle}{ 2 \langle D (\tau) \rangle }} =  \sqrt{ \frac{\theta}{\Omega (\tau)}}.
\EE
This measure is again universal, because it is directly related to the universal function $\Omega (\tau)$. For trait evolution near neutrality, the predictability decays to a small equilibrium value $\mathcal P_0 \approx \sqrt{\theta}$. But for sufficiently weak mutation rates and even moderate stabilizing strength, 
 the equilibrium value $\mathcal P_\stab = \sqrt{\theta / \Omega_\stab (c)}$ 
can be of order one: Stabilizing selection generates trait predictability over macro-evolutionary periods. Even if directional selection displaces the trait optimum $E^*$ to a new value that is common to the ensemble of populations, the trait distribution in each of the single evolved populations remains predictive of the ensemble trait repertoire at the same point in time. In a stochastic fitness seascape, lineage-specific changes of trait optimum  will reduce the predictability in the adaptive regime, $\mathcal P (\tau) = \mathcal P_\stab [1 - \Phi (\tau) + O(\tau^2)]$, but it can still retain values of order one over macro-evolutionary periods~\cite{Held13}. 
We conclude that the recent experimental observations of mutational stochasticity coupled with predictability at the level of molecular functions can be explained in a natural way, if we assume that many of these functions involve a complex quantitative trait.

\section*{Towards evolutionary systems biology}

We have discussed stabilizing selection as a mechanism that generates universal features of constraint and adaptation for a complex molecular trait. The same mechanism operates at different levels of molecular evolution. In a metabolic or regulatory pathway, for example, stabilizing selection on the function of a pathway generates evolutionary constraint on its output. This constraint is universal in the sense that it does not depend on fine-tuned levels and activities of the pathway components. Again, the reason for universality is that changes in one pathway component tend to be buffered by compensatory changes in other components. These compensations can be statistical or systematic, that is, generated by feedback loops in the pathway organization. Universality and predictability of pathway output emerge primarily in complex, higher-level pathways, which have multiple compensatory channels. This suggests the hierarchy of molecular functions  is reflected by an evolutionary hierarchy: universality and predictability increase, while stochasticity decreases with increasing level of complexity. This scale-dependence has been observed in the experiment of ref.~\cite{Tenaillon2012}: the level of predictability systematically increases with the level of genomic organization, ranging from individual mutations to genes, operons, and larger functional units. The increase of universality and predictability with scale may be limited by pleiotropy: lower-level system components are shared between different higher-level components~\cite{Maslov09}, which implies they are under additional evolutionary constraint.

Metabolic and regulatory networks, feedback loops, buffering, and stability are well-known topics of systems biology. Given the dramatic increase of evolutionary data at different molecular levels, they may also become elements of a quantitative evolutionary picture of molecular and cellular functions. As we have argued, this picture will involve hierarchies of nested quantitative traits with multiple fitness interactions, leading to scale-dependent universality and predictability. Making it quantitative will link systems biology to a new generation of molecular quantitative genetics, which remains largely to be developed. 

\section*{Acknowledgements}
We are thankful for stimulating discussions with  Shamil Sunyaev and  Jakub Otwinowski.  This work is supported by 
James S McDonnell Foundation  $21^{st}$ century science initiative-postdoctoral program in
complexity science/complex systems (Armita Nourmohammad) and by Deutsche Forschungsgemeinschaft grant SFB
680.

\newpage{}

\end{document}